\documentclass[manuscript]{acmart}
\AtBeginDocument{%
  }

\usepackage{tabularx}
\usepackage{array}

\begin{document}

%%
%% The "title" command has an optional parameter,
%% allowing the author to define a "short title" to be used in page headers.
\title{Is It Possible to Make Chatbots Virtuous? Investigating a Virtue-Based Design Methodology Applied to LLMs}

%%
%% The "author" command and its associated commands are used to define
%% the authors and their affiliations.
%% Of note is the shared affiliation of the first two authors, and the
%% "authornote" and "authornotemark" commands
%% used to denote shared contribution to the research.
\author{Matthew P. Lad}
\email{mlad@nd.edu}
%\orcid{}
%\affiliation{%
%  \institution{Institute for Clarity in Documentation}
%  \city{Dublin}
%  \state{Ohio}
%  \country{USA}
%}

\author{Louisa Conwill}
%\affiliation{%
%  \institution{The Th{\o}rv{\"a}ld Group}
%  \city{Hekla}
%  \country{Iceland}}
\email{lconwill@nd.edu}

\author{Megan Levis Scheirer}
%\affiliation{%
%  \institution{Inria Paris-Rocquencourt}
%  \city{Rocquencourt}
%  \country{France}
%}
\email{mlevis@nd.edu}

%%
%% By default, the full list of authors will be used in the page
%% headers. Often, this list is too long, and will overlap
%% other information printed in the page headers. This command allows
%% the author to define a more concise list
%% of authors' names for this purpose.
\renewcommand{\shortauthors}{Lad et al.}

%%
%% The abstract is a short summary of the work to be presented in the
%% article.
\begin{abstract}
With the rapid growth of Large Language Models (LLMs), criticism of their societal impact has also grown. Work in Responsible AI (RAI) has focused on the development of AI systems aimed at reducing harm. Responding to RAI’s criticisms and the need to bring the wisdom traditions into HCI, we apply Conwill et al.'s Virtue-Guided Technology Design method to LLMs. We cataloged new ethical design patterns for LLMs and evaluated them through interviews with technologists. Participants valued that the patterns provided more accuracy and robustness, better safety, new research opportunities, increased access and control, and reduced waste. Their concerns were that the patterns could be vulnerable to jailbreaking, were generalizing models too widely, and had potential implementation issues. Overall, participants reacted positively while also acknowledging the tradeoffs involved in ethical LLM design.

\end{abstract}

%%
%% The code below is generated by the tool at http://dl.acm.org/ccs.cfm.
%% Please copy and paste the code instead of the example below.
%%


%%
%% Keywords. The author(s) should pick words that accurately describe
%% the work being presented. Separate the keywords with commas.
%% A "teaser" image appears between the author and affiliation
%% information and the body of the document, and typically spans the
%% page.

%\received{20 February 2007}
%\received[revised]{12 March 2009}
%\received[accepted]{5 June 2009}

%%
%% This command processes the author and affiliation and title
%% information and builds the first part of the formatted document.
\maketitle

\section{Introduction and Related Work}
Large language models (LLMs) have exploded in popularity since the release of ChatGPT in November 2022~\cite{Marr_2023}. While they provide many benefits for productivity~\cite{Shone_2025}, they have also caused social and ethical concerns. These include environmental concerns~\cite{bender2021dangers, strubell2019energy, inie2025co2stly}, concerns with bias~\cite{bender2021dangers, gupta2025lawyers} and other harmful outputs including encouragement of eating disorders~\cite{choi2025private}, suicide~\cite{Kuznia_Gordon_Lavandera_2025}, and psychosis~\cite{fieldhouse2023can}, concerns with deskilling~\cite{Plaisance_2025}, and concerns about false information spread through LLM hallucinations~\cite{IBM_2025_hallucination}.

Responsible AI (RAI) principles have been proposed to guide the design, development, and deployment of AI systems, including LLMs, in ways that reduce harm. These principles include explainability, fairness, robustness, transparency, privacy, reliability, accountability, and inclusiveness~\cite{Stryker_2025, microsoft_responsible}. Prior work in HCI, however, has highlighted that while many of the current approaches to RAI focus on fixing harmful systems after they have been developed or deployed, there is a growing consensus that ethical concerns should be addressed in the early stages of development~\cite{saxena2025ai}. 

While RAI approaches remain critically important for building ethical AI, considering human values beyond RAI can provide a broader sense of how AI is impacting human flourishing. In particular, considering religious values and ancient wisdom in HCI and design can provide rich insights about the human experience~\cite{naqshbandi2022making}. 

Thus, we propose ethical designs for LLMs that are both addressed in the early stages of development, rather than being post-hoc fixes, and embody religious values. To do so, we employ the ``Virtue-Guided Technology Design'' approach published at CHI 2025~\cite{conwill2025}. In this paper, Conwill et al. put forth an approach for generating ethical design patterns that embody virtues from a chosen religious or philosophical tradition, and proposed seven design patterns for social media that embody the virtues of Catholic Social Teaching (CST) as a case study of their approach. We build on the work of Conwill et al., generating and evaluating design patterns for LLMs that embody the virtues of CST.

Through this new work, we contribute five design patterns to guide the development of LLMs in greater accordance with human flourishing. We also aim to spark conversation about using virtue ethics for LLM design.

\section{Methodology}
We follow the tripartite conceptual, technical, and empirical approach for documenting and evaluating virtue-oriented design patterns as outlined in Conwill et al.~\cite{conwill2025} Their first step is to choose the virtue ethics tradition to draw from and the technology to be built. For simplicity, we use the same principles of CST --- \textit{life and dignity of the human person; call to family, community, and participation; option for the poor and vulnerable; solidarity; subsidiarity;} and \textit{care of God's creation} --- that Conwill et al. used for their virtue tradition, and we of course are documenting design patterns for LLMs.

\begin{figure}
    \centering
    \includegraphics[width=\linewidth]{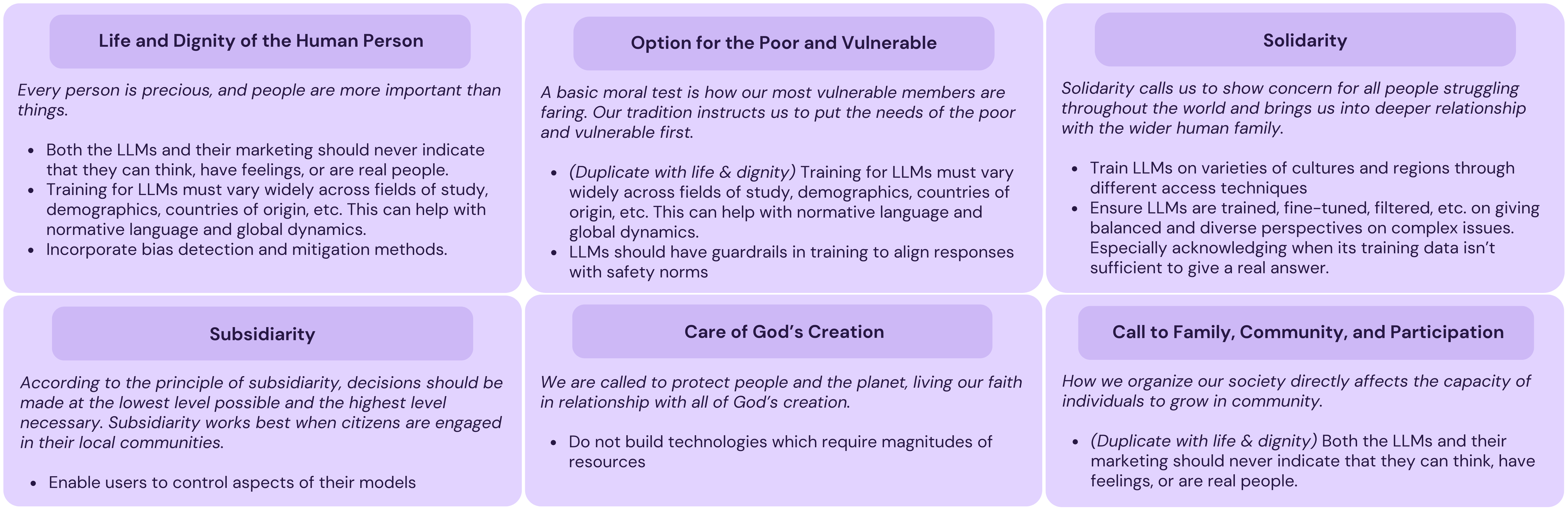}
    \caption{A summary of the conceptual inquiry, detailing how the six principles of Catholic Social Teaching apply to LLMs. For brevity we only show the points from the conceptual inquiry that ended up inspiring the five design patterns in this paper.}
    \label{fig:conceptual}
    \vspace{-15pt}
\end{figure}

In the \textbf{conceptual inquiry} we considered how our chosen virtues apply in the context of LLMs. For each virtue, we considered its definition from official Vatican sources and then reflected on various LLM use cases to determine how this virtue translates into the context of LLMs. The first author did an initial pass of this reflection; afterwards, the other authors contributed their thoughts on the first draft to make the final conceptual inquiry as robust as possible. A summary of the conceptual inquiry can be seen in figure \ref{fig:conceptual}.

\begin{figure}
    \centering
    \includegraphics[width=\linewidth]{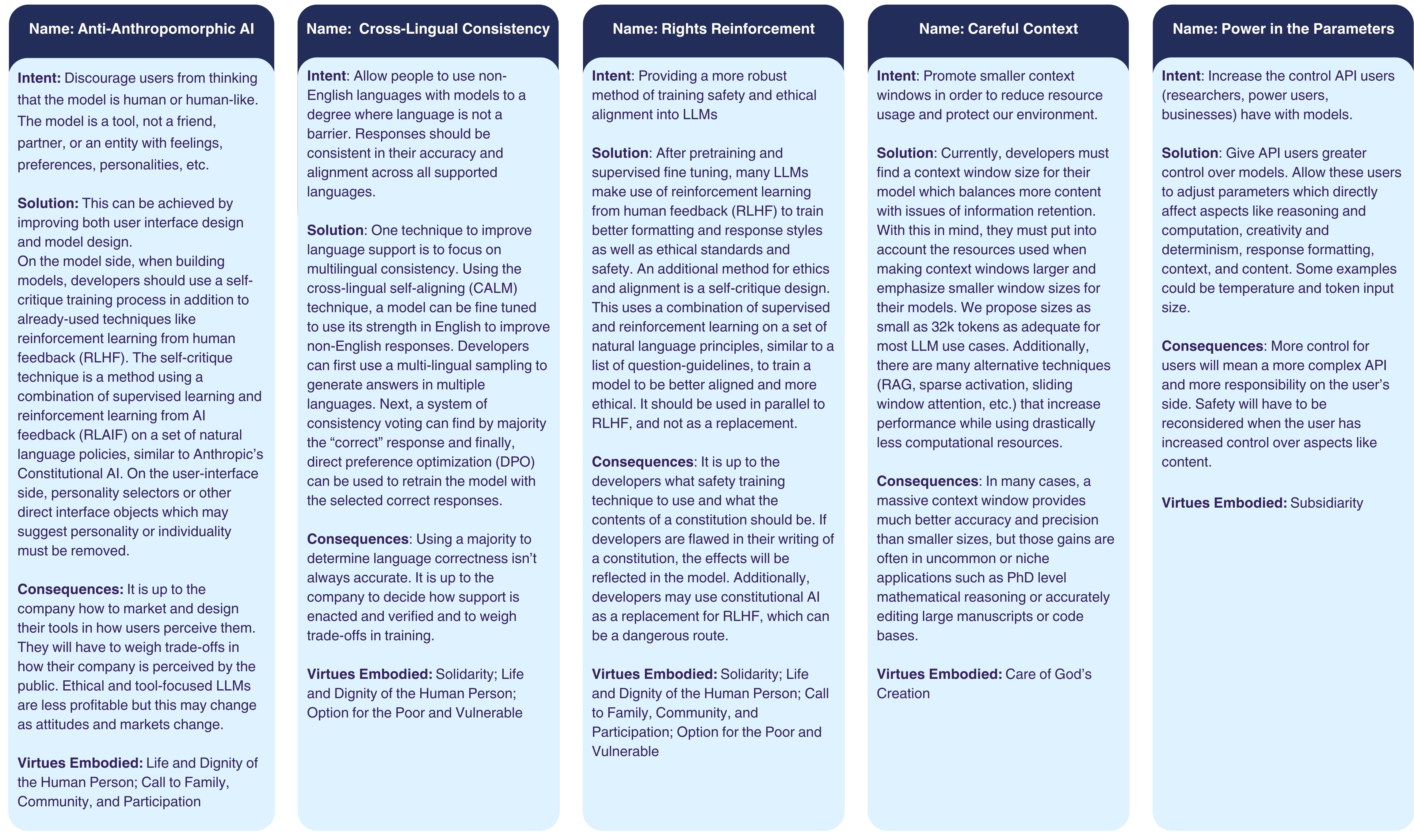}
    \caption{The five design patterns for virtuous LLMs that we documented through the Virtue-Guided Technology Design process.}
    \label{fig:design-patterns}
    \vspace{-25pt}
\end{figure}

In the \textbf{technical inquiry}, we identified already-existing LLMs and analyzed their technical features in light of the conceptual inquiry, documenting features that either uphold or violate our chosen virtues. We then cataloged design patterns based on patterns of upholding and violating features. After the first author generated 7 initial patterns, we held iterative group discussions with all authors plus one external reviewer to select and further refine 5 design patterns that had the best potential for meaningful discussion. The design patterns can be seen in figure \ref{fig:design-patterns}.

In the \textbf{empirical inquiry}, we conducted an IRB-approved interview study with technology researchers and industry practitioners to evaluate the design patterns. The interviews took a semi-structured format and entailed presenting participants with each design pattern (the consequences and virtues embodied were removed from the patterns to not bias responses) and asking them to describe how they thought the pattern would have a positive impact on LLMs if employed, a negative impact on LLMs if employed, and if they thought the pattern would have a net positive, net negative, or ambivalent impact overall. The full interview protocol is in the Appendix.

\textbf{Participant Recruitment and Demographics:} We recruited a small pilot group of 13 participants. Similarly to Conwill et al., participants were required to be at least 18 years of age and be a developer, researcher, designer, or graduate student with a background in computer science, software engineering, human-computer interaction, or related discipline. This technical background was required for adequate understanding of LLMs and better ability to evaluate design patterns. Participants were recruited through word of mouth and university communication channels such as relevant Slack teams. Participation was on a volunteer basis. Full participant demographics are in the Appendix.

\textbf{Analysis Approach:} Similar to Conwill et al., we used reflexive thematic analysis~\cite{braun2019reflecting} to analyze the responses, inductively coding for perspectives on the pros and cons of each design pattern. The first and second authors independently coded each transcript. Then, the entire research team worked together to discuss, debate, and synthesize the codes, searching for patterns of values and concerns and combining codes into overarching themes. This resulted in five themes for the aspects of the patterns participants valued, and five themes for the concerns with the patterns.

\textit{Positionality Statement:} All authors of this paper are white, American, and Roman Catholic. The first author is male and the second and third authors are female.

\section{Results}
We present a favorability rating for each design pattern, as well as the overarching values and concerns.

\begin{figure}
    \centering
    \includegraphics[width=\linewidth]{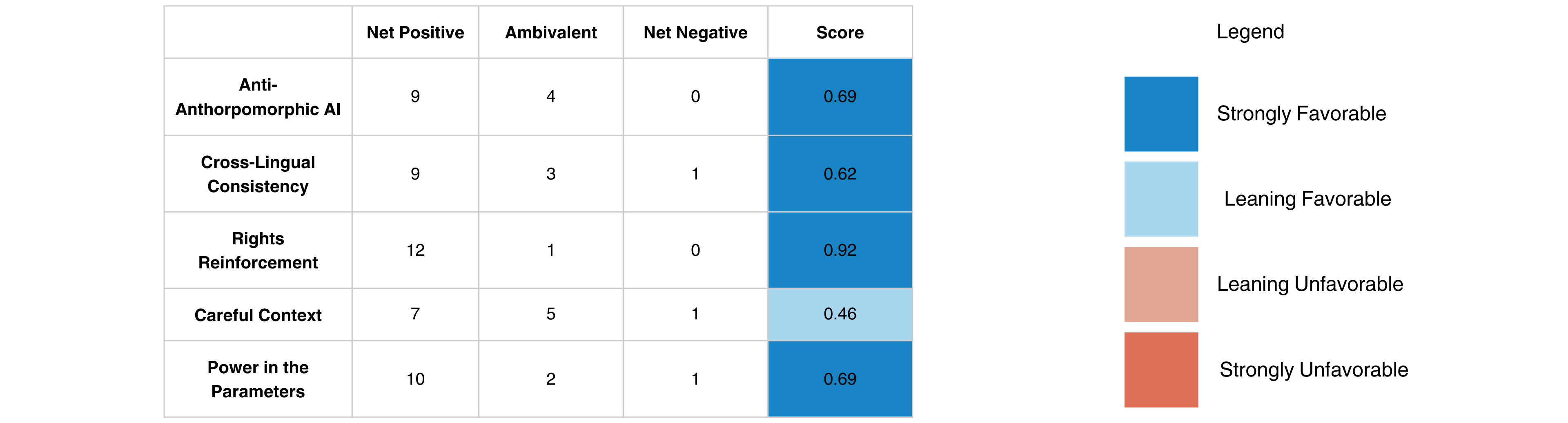}
    \vspace{-20pt}
    \caption{Total numbers of participants who described each pattern as a net positive, net negative, or ambivalent, and the score
describing if the pattern was overall seen as strongly favorable, leaning favorable, leaning unfavorable, or strongly unfavorable
by participants. All design patterns were favorable, and all patterns were strongly favorable except for Careful Context.}
    \label{fig:net-pos-neg-ambv}
    \vspace{-10pt}
\end{figure}

The favorability ratings are calculated using the same method as Conwill et al.: subtracting the number of net negative responses from the number of net positive responses and dividing that by the number of participants to determine a score from -1 to 1, where -1 indicates unanimous disfavor, and 1 indicates unanimous favor. The favorability scores can be seen in figure \ref{fig:net-pos-neg-ambv}. Most patterns were strongly favorable, and Careful Context was leaning favorable.

\begin{figure}
    \centering
    \includegraphics[width=\linewidth]{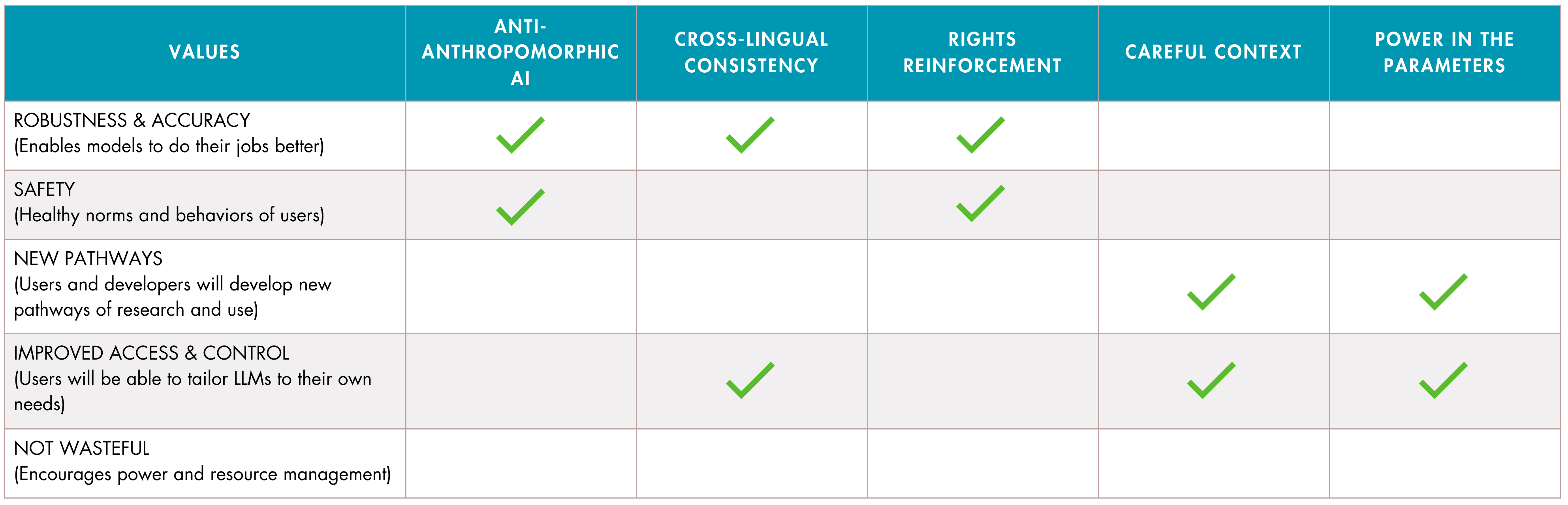}
    \caption{Overarching themes for what participants valued in corresponding proposed design patterns (green check).}
    \label{fig:values}
    \vspace{-10pt}
\end{figure}

The participant-identified values for the design patterns are \textbf{robustness \& accuracy}, \textbf{safety}, \textbf{new pathways}, \textbf{improved access \& control}, and \textbf{not wasteful.} Figure \ref{fig:values} shows which values were mentioned for each design pattern.

\textbf{Robustness \& accuracy} emphasizes any potential for models to provide better responses. One participant thought the Cross-Lingual Consistency Pattern \textit{``would...potentially help models with...correct responses in general.''} An additional aspect mentioned was the need for objective and less emotionally-charged responses. P2 spoke of it as a decrease in human tendencies, saying \textit{``... [LLMs] can be a little bit too fawning and positive about ideas. I think removing some of the human-like tendencies may have the result of giving...better advice.''}

\textbf{Safety} is described as healthy behaviors and norms for interacting with LLM chatbots. This includes chatbots not being a \textit{``replacement of a person''} (P8), not reinforcing negative ideas that could lead to unhealthy behaviors like suicide (P6), and being more transparent with responses, emphasizing that the information is coming from \textit{``an AI model and not from an expert in the field''} (P1). Multiple participants mentioned the danger of replacing therapists with chatbots. 
 
\textbf{New pathways} refers to encouraging development of new avenues of research, improved workflow, and the need for user education. Participants mentioned that the Careful Context pattern could \textit{``encourage more research and use of alternative techniques''} (P2) and \textit{``could force [users] to scope problems correctly''} (P8). Training on smaller context windows could even give LLMs a more positive \textit{``image''} by improving their environmental impact (P9). The Power in the Parameters pattern could help \textit{``people who know what they're doing''} harness more power from LLMs (P1). 

\textbf{Improved access \& control} refers to a diverse set of users being able to customize LLMs for their own purposes. This includes expanding LLMs to more languages for increased accessibility: \textit{``if there's someone...who's able to articulate their thoughts better in a different language, then they would be able to use these LLMs to a better degree''} (P9). In terms of customization, P2 said the Power in the Parameters pattern allows \textit{``a better efficacy from the model by adjusting things like temperature, token input size, they can tailor it better to their specific scenario.''}

\textbf{Not wasteful} refers to a design pattern saving power, resources, and being generally better for the environment, as well as saving time (e.g. faster inference). Participants said about the Careful Context pattern that \textit{``it would definitely improve the amount of power that [LLMs] suck from our infrastructure...and it might even make them...a little faster''} (P1) especially given that \textit{``models are generally required to have tons and tons of compute''} (P7). P8 also noted that the Careful Context pattern would \textit{``force [them] to scope...problem[s] better''}, leading to more efficient solutions.

\begin{figure}
    \centering
    \includegraphics[width=\linewidth]{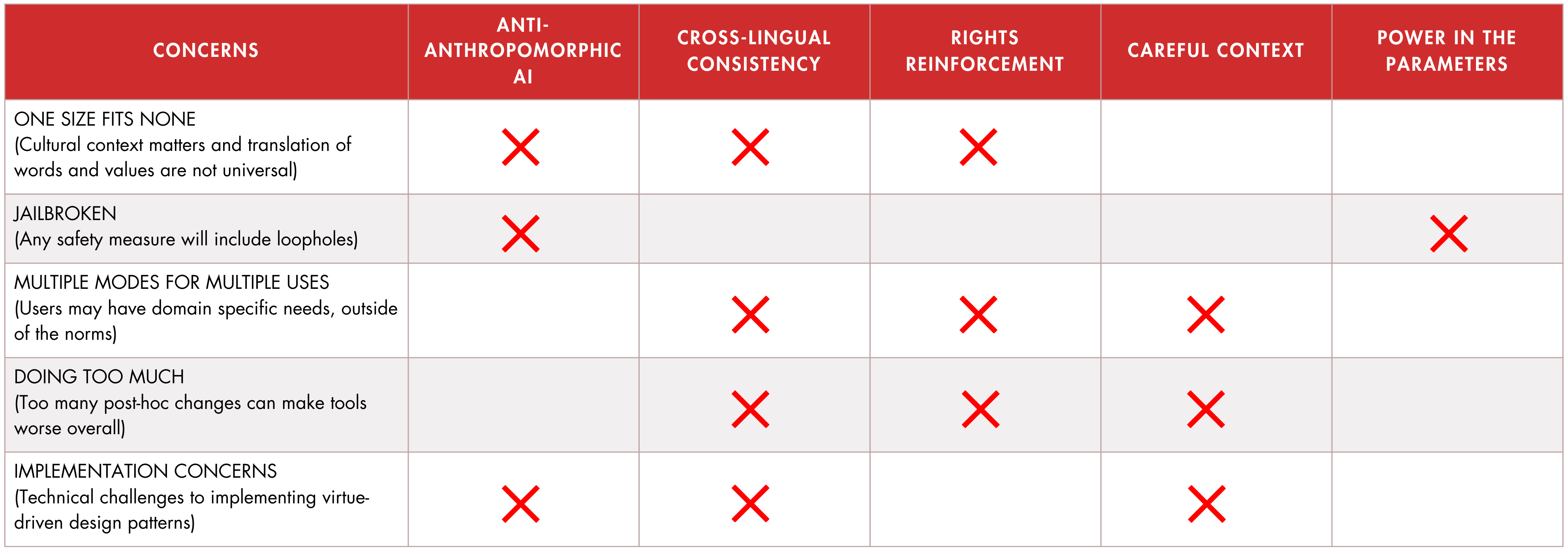}
    \caption{Overarching themes for participants' concerns with indicated corresponding proposed design patterns (red X).}
    \label{fig:concerns}
    \vspace{-21pt}
\end{figure}

The concerns for the design patterns are \textbf{one size fits none}, \textbf{jailbroken}, \textbf{multiple modes for multiple uses}, \textbf{doing too much}, and \textbf{implementation concerns}. Figure \ref{fig:concerns} shows which concerns were mentioned for which patterns.

\textbf{One size fits none} refers to concerns that the design patterns will reinforce lingual or cultural norms from a dominant culture. This includes a risk of incorrect translations or cultural norms getting baked into the responses: \textit{``I think it's important to think about localization and in this case, cultural differences...if you want to improve non-English responses using English data, you're...using English, or...probably American culture as the basis for those answers...someone in a different culture who's using the LLM in their language may find American cultural markers in the responses''} (P2). This theme also encompasses concerns with ethical pluralism and alignment. Not only may \textit{``my set of ethics might be different from someone else's''} (P8) but also encoding a particular ethical standard may force behaviors that not all users align with:  \textit{``by putting these [ethical] guardrails, you're...influencing the way people are going to use these systems''} (P8).

\textbf{Jailbroken} refers to the concern that safety measures or design patterns with a generally positive intent will have loopholes. Participants expressed concerns with Power in the Parameters that \textit{``when giving users more control, there's more potential for abuse...some of these parameters might...allow the users to more easily jailbreak the model or avoid some of the guardrails''} (P6). In relation to the Anti-Anthropomorphic AI pattern, participants expressed concerns about the pattern not working as intended due to the human inevitability of anthropomorphization, or on the flip side, that people may trust AI systems more than humans: \textit{``maybe emphasizing the fact that it's this computing system that's producing this output for you might lead people to trust it even more''} (P6).

\textbf{Multiple modes for multiple uses} refers to a design pattern not allowing for users' particular domain-specific needs. One such example is the Careful Context Pattern not allowing for large enough context windows in legitimate use cases: \textit{``if you...[need a larger context window and]...don't have an option to increase it, that's a problem''} (P10). Additionally, the Rights Reinforcement pattern could inadvertently block valid and ethical use cases: \textit{``if I were to ask ChatGPT right now about some darker situations, it's gonna come back and say, hey, that's probably not good to be thinking about. But if I'm a police investigator, I may need to know about some of those darker situations in a context of investigative work''} (P5). Finally, the Cross-Lingual Consistency Pattern may need human, domain-specific verification for critical use cases: \textit{``using it in [medical or legal] settings... with non English languages, where...exact human verified wording is compulsory...you need domain experts or human translator[s]''} (P12).

\textbf{Doing too much} refers to the concern that the design patterns will cause worse accuracy or user experience. For example, smaller context windows could require users to \textit{``give even more prompts to do stuff''} (P3) or could cause hallucinations at \textit{``the upper bounds of the window size''} (P13). Participants also expressed concerns with the design patterns over-promising and under-delivering in accuracy or ethical standards, including \textit{``[companies] doing something that they believe is improving safety and ethical alignment that is...resulting in answers that are not truthful''} (P2).

\textbf{Implementation concerns} refers to a belief that there are technical limitations to implementing the design pattern in the intended way. Participants noted technical feasibility, increased compute time, dilution, data privacy, and unintended environmental impacts as potential implementation concerns across the design patterns.

\section{Discussion and Conclusion}
In both the participant values and concerns, participants expressed a desire to move away from general-purpose LLMs towards models that are smaller and customized for particular purposes. Namely, the improved access \& control value relates to customizability and not wasteful relates to smaller models. The concerns one size fits none and multiple modes for multiple uses emphasize the need to create specialized LLMs for different use cases. Considering these values and concerns together, they point to the design implication that smaller models that are specialized for particular use cases are more desirable than the generalized LLMs that are common today.

Most participants rated the design patterns as having a positive impact on LLMs, yet they frequently mentioned that each design pattern had trade-offs rather than being unequivocally positive. Participants also frequently expressed that even if the ethical guidelines proposed in the design patterns weren’t perfect, some ethical guidelines are better than none at all. Taken together, this indicates that while ethics can be tricky to implement, LLM users value some implementation of ethics, even if not perfect. Given that Rights Reinforcement was the most favored proposed design pattern, attempting to implement this pattern would be a next step in this research to reveal whether or not the tradeoffs and implementation concerns hold.

Applying Virtue-Guided Technology Design to LLMs surfaced challenges with the method that Conwill et al. did not experience when generating social media designs. First, LLMs feature a more universal design, usually a chatbot. Social media platforms have a wider range of designs, making it easier to generate diverse design patterns. Second, while social media is technically mature, LLMs are newer and thus still under algorithmic development. This led to a number of participants voicing “implementation concerns,” unsure if the algorithms in the proposed design patterns would work. This points to the importance of further algorithmic development for LLMs, which can turn ethical ideals into working implementations. 

%% The acknowledgments section is defined using the "acks" environment
%% (and NOT an unnumbered section). This ensures the proper
%% identification of the section in the article metadata, and the
%% consistent spelling of the heading.
%\begin{acks}
%to-do
%\end{acks}

%%
%% The next two lines define the bibliography style to be used, and
%% the bibliography file.
\bibliographystyle{ACM-Reference-Format}
\bibliography{sample-base}

%%
%% If your work has an appendix, this is the place to put it.
\appendix

\end{document}

\title{Appendix for Is It Possible to Make Chatbots Virtuous? Investigating a Virtue-Based Design Methodology Applied to LLMs}

\author{Matthew P. Lad}
\email{mlad@nd.edu}

\author{Louisa Conwill}
\email{lconwill@nd.edu}

\author{Megan Levis Scheirer}
\email{mlevis@nd.edu}

\renewcommand{\shortauthors}{Lad et al.}

\maketitle

\section{Participant Demographics}

\begin{table*}[htb]
\caption{\textbf{Participant Demographics}}
\label{tab:participant_demographics}
\renewcommand{\arraystretch}{1.2}
\begin{tabularx}{\textwidth}{|c|c|c|c|X|X|}
\hline
\textbf{ID\#} &
\textbf{Gender} &
\textbf{Race/Ethnicity} &
\textbf{Religious Affiliation} &
\textbf{Educational Experience} &
\textbf{Technical Work Experience} \\
\hline
P1 & Male & No answer & Catholic &
BS in CS &
Cybersecurity internship, IT work, medical device repair \\
\hline
P2 & Female & White & Catholic &
BS in CS, Current PhD Student in CS &
Software engineering internships \\
\hline
P3 & Male & White & Catholic &
BS in CS, Current PhD Student in CS, Current MS Student in Theology &
Software engineering internships \\
\hline
P4 & Male & White & Catholic &
BS in CS, Current MS Student in CS &
None reported \\
\hline
P5 & Male & White & Protestant (Presbyterian) &
BS in CS &
Data science and analysis \\
\hline
P6 & Male & White & Catholic &
BS/MS in CS, Current PhD Student &
Software engineering \\
\hline
P7 & Male & Hispanic/Latino & Catholic &
BS and PhD in CS &
Postdoctoral research in CS \\
\hline
P8 & Male & Hispanic/Latino & Catholic &
BS in CS, Current PhD Student in CS &
Software engineering \\
\hline
P9 & Male & White & Catholic &
Current BS Student in CS &
Software engineering internship \\
\hline
P10 & Male & White & Reformed (Protestant) &
BS in CS, Current PhD Student in CS &
Undergraduate research in CS \\
\hline
P11 & Male & White & Catholic &
BS in CS &
IT support \\
\hline
P12 & Male & South Asian & Catholic &
BS and MS in CS, Current PhD Student in CS &
Machine learning internship \\
\hline
P13 & Male & White & Catholic &
BS in Math, MS in Teaching, Current PhD Student in Engineering &
High school teaching in math, CS, and robotics \\
\hline
\end{tabularx}
\end{table*}

Our participant demographics are described in detail in table \ref{tab:participant_demographics}.

\section{Interview Protocol}

We conducted semi-structured interviews based on this set of base questions. If a participant did not have a solid understanding of what a design pattern is, the following was read to them:

In software engineering, a design pattern is a tool for sharing design knowledge, especially about designs that solve commonly-occurring problems. Design patterns encapsulate the description of a common design problem, its context, and the solution suggested by the pattern. In this interview, we will present to you various design patterns that have been created for large language models (LLMs). The goal of these patterns is to direct the development of LLMs such that they are good for the human person and avoid many of the pitfalls we see with artificial intelligence (AI) today.

\subsection{Interview Questions}

The following questions were asked of each participant before showing them the design patterns.

\textbf{Pre-Questions (Basic Demographic Information): }
All of these questions are optional to disclose
\begin{itemize}
    \item What is your gender?
    \item How would you describe your race/ethnicity?
    \item What is your religious background?
\end{itemize}

\noindent\textbf{Question 1:} What is your educational and professional background, especially with regards to technology?

\noindent\textbf{Question 2:} In your day to day work, do you make or have you made any design decisions? (for example at the software level or the user experience level)

\noindent\textbf{Question 3:} How do you typically make design decisions?

\noindent\textbf{Question 4:} What do you prioritize when making design decisions?

\noindent\textbf{Question 5:} What is your experience with using design patterns?

\noindent\textbf{Question 6:} In your experience using design patterns, how have you found them to be helpful (if at all)?

\noindent The participants were shown each design pattern and asked the following four questions about each one. They were also asked follow-up questions if necessary.

\noindent\textbf{Question 7:} What clarification questions do you have (if any) about how the pattern works?

\noindent\textbf{Question 8:} In what ways (if any) do you think this design pattern would improve large language models?

\noindent\textbf{Question 9:} In what ways (if any) do you think this design pattern would not improve large language models?

\noindent\textbf{Question 10:} Do you think this pattern will have more of a net positive, net negative, or an ambivalent impact on large language models?